\documentclass[manuscript,nonacm]{acmart}

\usepackage{rotating}
\usepackage{type1cm} %
\usepackage{graphicx} %
\usepackage{xspace} %
\usepackage{balance} %
\usepackage{booktabs} %
\usepackage{multirow} %
\usepackage[font={bf}, tableposition=top]{caption} %
\usepackage{bold-extra} %
\usepackage{siunitx} %
\usepackage[vlined,linesnumbered,ruled,noend]{algorithm2e} %
\usepackage{microtype} %
\usepackage{xfrac} %
\usepackage{mathtools} %
\PassOptionsToPackage{hyphens}{url} %
\PassOptionsToPackage{bookmarks, pdftex, colorlinks=true, pagebackref=true, backref=page}{hyperref} %
\usepackage{cleveref} %
\PassOptionsToPackage{square,numbers,backref=false}{natbib} %
\usepackage[hyperpageref]{backref} %
\usepackage{hyphenat} %
\usepackage{subcaption} %
\usepackage[export]{adjustbox} %
\usepackage[show]{chato-notes}
\usepackage{placeins}

\renewcommand*{\backref}[1]{}
\renewcommand*{\backrefalt}[4]{}

\graphicspath{{fig/}}

\renewcommand*\backref[1]{\ifx#1\relax \else (Cited on #1) \fi}

\newcommand\Tf{\rule{0pt}{2.8ex}}

\title{Conditional Publics: Shared Events and Divergent Meanings in the European Twitter Debate on the Ukraine War}

\author{Corrado Monti}
\orcid{0000-0001-6846-5718}
\affiliation{%
  \institution{CENTAI}
  \city{Turin}
  \country{Italy}
}
\email{me@corradomonti.com}

\author{Arthur Capozzi}
\orcid{0000-0002-1996-1800}
\affiliation{%
  \institution{ETH Zürich}
  \city{Zürich}
  \country{Switzerland}
}
\email{arthur.capozzi@gess.ethz.ch}

\author{Yelena Mejova}
\orcid{0000-0001-5560-4109}
\affiliation{%
  \institution{ISI Foundation}
  \city{Turin}
  \country{Italy}
}
\email{yelenamejova@acm.org}

\author{Gianmarco De Francisci Morales}
\orcid{0000-0002-2415-494X}
\affiliation{%
  \institution{CENTAI}
  \city{Turin}
  \country{Italy}
}
\email{gdfm@acm.org}

\begin{abstract}
How do European publics debate a geopolitical crisis on social media, and do they inhabit a shared informational reality?
We analyze over 38 million geolocated tweets from 20 European countries during the first eight months of the Russian invasion of Ukraine.
Using retweet community detection and stance annotation across six issues, we identify `hawkish' and `doveish' opinion clusters present within almost every country studied.
We find that structural polarization is driven not by radicalization, but by the exit of casual users.
Crucially, whether opposing sides orient to the same events depends on the issue.
On pragmatist issues, both sides react to the same high-profile events, forming an agonistic public sphere.
Instead, on interpretive issues, they operate as affective publics and counterpublics constructing divergent meanings.
We propose \emph{conditional publics} to describe formations whose relational structure, sharing or fracturing a referential frame, depends on the epistemic character of the debated issue.
\end{abstract}

\keywords{Public Sphere, Social Media, Ukraine War, Polarization}

\begin{document}

\maketitle

\section{Introduction}
\label{sec:intro}

An immense human tragedy, the 2022 Russian invasion of Ukraine was also a turning point for European politics, identity, and security order~\cite{duck2025introduction}.
It was widely framed by European elites as a revolutionary moment of transformation, as illustrated by Olaf Scholz's \emph{Zeitenwende} (``era shift'') speech.
The war undoubtedly contributed to reshaping European narratives, accelerating a realignment of European identity along the fault lines of its relationship with Russia~\cite{makarychev2014russiaAndEU}.
At the same time, the Ukraine war is one of the first prominent examples of a large-scale interstate conflict fought on digital media battlefields~\cite{mejova2025narratives}.
On social media, information and narratives are constantly contested, with different ideological camps actively building their own information ecosystems and talking points in competing attention economies~\cite{abidin2021networked}.

The European social media debate around the war thus offers an opportunity to study how publics construct political meaning during a crisis.
Some parts of European identity have come into existence precisely through the antagonism with Russia, making this debate a climax in a long history of interdependent yet conflictual relations between the two worlds~\cite{makarychev2014russiaAndEU}.
At stake is a foundational question for the study of digital public spheres:
whether socially mediated debate around a shared crisis produces a common informational reality, or whether opposing publics construct entirely separate universes of relevant events, actors, and meanings.

In this work, we study how these phenomena interact from a particular vantage point: the social media debate in Europe on the Ukraine war.
We choose Twitter (now X) as our object of study because of its prominent role in then-current news commentary, especially among journalists, pundits, and political actors.
Social media platforms, such as Twitter, provide an ideal setting for observing the processes through which opinions shift as audiences select, amplify, and challenge interpretations of events.
Specifically, we ask:

\begin{itemize}
\item \textbf{RQ1:} How does the composition of active users interact with the dynamics of polarization in the European debate after the Russian invasion?
(\S\ref{sec:polarization_over_time})
\item \textbf{RQ2:} How do these dynamics and temporal patterns of attention relate to the different stances? (\S\ref{sec:opinion_polarization})
\item \textbf{RQ3:} To what extent are these attention dynamics coordinated at the European level, as opposed to being nationally fragmented? (\S\ref{sec:Opinion_expression_synchronizes_across_Europe})
\item \textbf{RQ4:} Do opposing sides react to the same external events,
or do they construct divergent attention horizons by selecting different events as salient,
and does this depend on the nature of the issue being discussed? (\S\ref{sec:intrinsic_vs_extrinsic_focus})
\end{itemize}

To answer these questions, we analyze a large-scale dataset of tweets (600M+) related to the Ukraine war collected between February and October 2022.
We find that, following an initial surge in tweet volume, public attention declines accompanied by an increase in structural polarization.
This trajectory reflects a selection effect: the user base becomes progressively dominated by long-term, highly engaged users
as more casual participants drop out.
Mapping the most polarizing tweets onto six issue axes, including responsibility for the conflict, sanctions, and the provision of weapons to Ukraine, we identify two major clusters of users that diverge both in opinion and in temporal activity.
On the whole, these factions attend to different events and amplify different viral content, constructing largely separate informational flows.
However, when both sides converge on the same events, the subject matter concerns \emph{pragmatist} issues: debates around specific institutional decisions and policy instruments.
On \emph{interpretive} issues such as responsibility and desired outcomes, the two sides foreground entirely different events and historical references, constructing divergent meanings from the same conflict.
We propose the concept of \emph{conditional publics} to account for these dynamics:
formations whose relational structure, specifically whether they share or fragment a common referential frame with opposing communities, depends on the epistemic character of the issue under debate~\cite{risse2010community,koopmans2010europeanization}.
In other words, the extent to which opposing groups share attention to the same events varies systematically with the type of issue under discussion.

\section{Conceptual Framework}
\label{sec:framework}

\subsection{Constructivist Perspective}

To properly characterize how European social media users have framed the war within the study of international relations, we adopt a constructivist framework.
Constructivism in international relations starts from the premise that international politics is not governed solely by material power or fixed interests (as emphasized by realism)~\cite{guzzini2013power}, but by socially constructed meanings, identities, and norms, 
produced through social interaction, discourse, institutional practices, and through public debate, including social media.
From this perspective, interests themselves are shaped by shared ideas about legitimacy, responsibility, security, and appropriate behavior~\cite{guzzini2013power}.

Social media debate on the Russia-Ukranian conflict has been ongoing online since the rise of the separatism in the Donbass region,
signified by hashtag \#SaveDonbassPeople \cite{noakes2005frames,makhortykh2015savedonbasspeople,makhortykh2017social}.
As the debate involves concepts such as who is responsible for the conflict and what is the definition of a desirable outcome, 
it is natural to adopt a constructivist lens.
Constructivist scholars such as
Andrey Makarychev showed how EU–Russia–Ukraine relations are structured by competing narratives and normative claims, rather than by material power calculations alone~\cite{makarychev2025tandf_pending}.
Sovereignty, Europeanness, legality, and moral authority function as symbolic markers of identity~
\cite{makarychev2013nonwestern_theory}, 
suggesting that the conflict is mediated through contested meanings.

\subsection{Interpretive and Pragmatist Issues}
\label{sec:interpretive-pragmatist}

We develop our analysis via a theoretically-informed Grounded Theory approach \cite{charmaz2017constructivist} whereby the data analysis is interspersed with theoretical construction and contextualization. 
After an iterative reading of the data and theoretical refinement, we developed a set of thematic axes to guide our analysis (which is described in depth in ~\Cref{sec:axis-labeling}).
These topical axes can be broadly categorized into two macroscopic areas corresponding to complementary epistemiological traditions: \emph{interpretive} and \emph{pragmatist}~\citep{goldkuhl2012pragmatism}.
Interpretive issues emphasize the meaning given to social constructs, where broad signifiers such as aggression, blame, and punishment are interpreted differently by different parties.
The interpretive axes focus on understanding the context, are more normative (exhibiting values), and concern how actors frame causality, goals, and Europe's role.
Conversely, pragmatist axes concern actions and their effects on the world, e.g., the means through which legislative and judiciary powers concretely respond to the war.
These axes are more prescriptive and concern whether specific means (weapons, sanctions) and judgments (war crimes attribution) are considered legitimate.
This distinction reflects a difference between issues structured primarily by normative interpretation and narrative framing and those anchored in externally verifiable events.
Pragmatist issues impose external referential constraints through discrete and widely-observable events, whereas interpretive issues lack such anchors and allow greater selectivity in the construction of relevance.

Regarding the interpretive axes, the issues we identify are about who is assigned responsibility for the war and what ideal outcome Europeans should hope for.
For the former, we word it simply as ``\emph{Which party is emphasized as responsible for the outbreak of the war in Ukraine?}''.
This question was asked by the European Council on Foreign Relations in a May 2022 survey, reporting a wide dispersion among the European countries' assigning the responsibility to either Russia or Ukraine, the EU, and the US (the last three grouped together) \cite{krastev_leonard_2022_peace}.
Further, the survey included value judgments on the future of the conflict, with the question \emph{``What is the most important outcome of the war?''} presenting two selections: \emph{``The most important thing is to stop the war as soon as possible, even if it means Ukraine giving control of areas to Russia''}, \emph{``The most important thing is to punish Russia for its aggression, even if it means that more Ukrainians are killed and displaced''}, plus \emph{``Neither of these''}, and \emph{``Don’t know''}. 
Those who selected ``stop the war'' option were considered to be in the ``Peace camp'', while those who chose the ``punish Russia'', in the ``Justice camp'' \cite{krastev_leonard_2022_peace}. 
A more recent 2024 version of the survey \cite{krastev_leonard_2024_sovereignty} includes a more concrete choice: \emph{``What should Europe do about the war in Ukraine?''} with possible actions being \emph{``Europe should push Ukraine towards negotiating a peace deal with Russia''} or \emph{``Europe should support Ukraine to fight to regain the territories occupied by Russia''}, which again showed a large variability across European nations.
Together, these questions exemplify the interpretation of past events and the values ascribed to future outcomes and actions. 

Regarding the pragmatist axes, instead, we identify three topics as divisive in the European debate and closely tied to the legitimation of concrete measures and judicial claims: the provision of weapons to Ukraine, the imposition of economic sanctions on Russia, and the attribution of war crimes.
For the first two, we operationalize these debates through the following questions, following the wording used by public surveys~\cite{eurobarometer_3053_2024,eupinions_war_vote_2024}: ``\emph{Should your government support Ukraine by delivering weapons?}'' and ``\emph{Do you agree with imposing economic sanctions on the Russian government, companies, and individuals?}'', which capture disagreement over the legitimacy and acceptability of coercive instruments. 
The third instrumental axis concerns the attribution of war crimes and is phrased as ``\emph{Did one of the two sides attack civilians, execute prisoners, or perpetrate other unlawful acts?}'', reflecting contestation over legal judgments and the internationally-accepted definition of war crimes.
Taken together, these pragmatist axes allow us to capture polarization over how the war should be confronted in practice, 
focusing on the European normative role~\cite{hoffmann2018russia} as analyzed by \citet{makarychev2014russia}. %
We give the complete description of questions and possible answers in \Cref{tab:issues}.

\subsection{Issue Polarization on Social Media}

Social media has long been theorized as a space for deliberation and identity formation, and has accordingly been proposed as a potential substrate for the emergence of a European transnational public sphere~\cite{risse2010community,koopmans2010europeanization}. 
However, the engagement-optimization logic embedded in platform algorithms structurally favors divisive and emotionally charged content, creating conditions that are fundamentally inhospitable to cross-cutting deliberation and consensus formation~\cite{rathje2021out,brady2020attentional}.

Twitter specifically has been shown to be prone to foster divisiveness, especially around controversial topics~\cite{garimella2018political, cinelli2021echo}. 
On this platform, the affordances of reposting (or retweeting) a post without providing additional commentary is seen as a signal of agreement, or endorsement. 
A network composed of users connected by one reposting another is called an endorsement network \cite{garimella2018quantifying}.
Endorsement networks have been used to discover ``communities'' of social media users (with ``communities'' used extremely loosely) that share stances on a topic.  %
These communities then provide the analytical units through which to assess to what extent the interactions are homophilic, indicating greater separation from the other discussants~\cite{sunstein2002law}.

\section{Data \& Methods}
\label{sec:datamethods}

\subsection{Data collection} 

The data used in this study was collected via the Twitter Streaming API\footnote{\url{https://developer.x.com/en/docs/tutorials/consuming-streaming-data}} using an extensive query of war-related keywords (following~\citep{mejova2025narratives}), which include ``Ukraine'' in over 50 languages.
The collection comprises \num{623945097} tweets posted between 27 February and 12 October 2022. 
The API provides tweets and their retweets at the time of their posting (in ``streaming'' fashion). 
The API also provides a language tag, computed by the platform on the content of the tweet, as well as a ``geo'' tag concerning the post's geographic location.\footnote{Twitter (2021b) Data Dictionary: Standard (version 1.1). Developer Platform. Available at: \url{https://developer.twitter.com/en/docs/twitter-api/v1/data-dictionary/object-model/tweet} (retrieved on \today)}
Unfortunately, the geographic information is almost never provided, thus we proceed to geo-locate the posts through users' Location field.

Following previous literature \cite{lenti2023global}, we map the user Location field to the geographic location database GeoNames.\footnote{\url{https://www.geonames.org}}
Because this field is entered by the user and not validated by the platform, we manually check the top 3000 most popular location matches to exclude those that are not locations (such as ``on the moon''). 
Note that we assume that the tweet is posted from the location that the user has disclosed. This approach was validated~\cite{paoletti2024political} by comparing the extracted locations to those specified in the ``geo'' field, resulting in 93.7\% accuracy.
This resulted in \num{184049190} tweets associated with a location.
Because our focus is on Europe, we considered the tweets from European countries having at least \num{10000} users in our data, plus those bordering Russia, resulting in a total of \num{20} countries.
Note that we do not include Russia and Ukraine in this analysis, as we are interested in the reaction to the war by countries not directly involved in the conflict, for interpretative clarity. 
Finally, for each country, we selected tweets in the language that is most used in that country, as per our dataset (using the ``lang'' tag provided by the Twitter API).
For 8 countries (including Germany, France, Italy and Spain) a national language was selected, while for the others, English resulted as the most used language 
(for instance, 72\% of the tweets from Finland were in English, and only 18\% in Finnish). 
The final dataset consists of \num{38044266} tweets, out of which \num{11122581} are original tweets, and the rest are retweets.

\subsection{Polarizing tweets}
\label{sec:polarizingtweets}

To study the dynamics of the debate, we identify a set of polarizing
tweets, that convey the most divisive opinions.
This design choice allows us to delineate the boundaries of the debate by focusing on a fundamental set of tweets.
To identify this set of tweets among the \num{38} million, we follow a data-driven approach.

First, we run a community detection algorithm on the retweet network of each country separately.
We use the Leiden algorithm~\cite{traag2019louvain} as it discovers the number of communities automatically from the data.
Technically, we run the algorithm on the largest weakly connected component of the retweet network after removing dangling nodes (with degree $1$).
For each country, we keep the largest communities by number of users until we reach a 75\% coverage of the network.
This process results in $3$ to $7$ communities per country, with an average of $4.35$.

Then, we extract the most representative tweets of each community as follows.
For each tweet retweeted in a community, we define its polarization score as $n_i - n_o$, where $n_i$ is the number of retweets within that community, and $n_o$ is the number of retweets outside.
We extract the 20 tweets with highest score for each country and community.
This method yields a dataset of \num{1242} unique tweets, which are responsible for \num{4958160} retweets  %
(3.75\% of total retweets): the most divisive with respect to the retweet network.

\subsection{Axis labeling}
\label{sec:axis-labeling}

All four authors of this work have read through all of the selected tweets and annotated them for the poster's stance 
on the issues described in \Cref{tab:issues}.
If a tweet does not express any of the coded stances on an issue, it is not annotated, and it receives a score of 0.
To ease the annotation, the content that was not in English was translated into English using Google Translate, and the authors referred to the original tweet (if one was still available) via the link in case the textual content was insufficient to make a decision.
Further, if additional information was necessary to make a decision, the annotators explored the news at the time of the tweet's posting to understand its context.
The annotation was performed separately by the four annotators, with occasional discussion around unclear cases. 

For the continued analysis of the tweets, for each axis, we aggregate the scores made by each of the four annotators by coding them to integers as indicated in \Cref{tab:issues} and averaging the scores.
For instance, a tweet that received ``Nato or Ukraine'' on the Responsibility issue from 3 annotators, and none from 1 annotator, would have a score of $0.75$ for this issue.
This way, we retain and average individual annotations rather than collapsing them into a single majority label.
This design choice reflects a perspectivist stance toward ground truth in subjective NLP tasks~\cite{ovesdotter-alm-2011-subjective,cabitza2023toward},
as disagreement among annotators may reflect genuine differences in interpretation rather than annotation noise or guideline failure~\cite{plank2022problem,uma2021learning}.
Majority voting, or using inter-annotator agreement as a quality measure, in our context would obscure pluralism~\cite{Frenda_2024} and contradict our constructivist framework.

\begin{table}[th] %
\caption{Questions around issues related to the Ukraine war, along with the possible answers and their binary encoding.
The data in \Cref{fig:communities_issues} shows that the six axes present correlations in opinion.
}
\begin{center}
\begin{tabular}{l p{0.45\linewidth} p{0.38\linewidth}}
\toprule
Title & Question & Possible answers (encoding) \\
\midrule
Responsibility & Which party is emphasized as responsible for the outbreak of the war in Ukraine? & %
    Nato or Ukraine (-1), \newline
    Russia (+1) \\ \addlinespace 
Outcome & What is the most important outcome of the war? & %
    Stop the war as soon as possible (-1), \newline
    Punish Russia for its aggression (+1) \\ \addlinespace  %
Act & What should Europe do about the war in Ukraine? & %
	Push Ukraine to negotiate (-1), \newline
    Support Ukraine in its fight (+1) \\ \addlinespace 
Weapons & Should your government support Ukraine by delivering weapons? & %
	No (-1), \newline
    Yes (+1) \\ \addlinespace 
Sanctions & Do you agree with imposing economic sanctions on Russian government, companies and individuals? & %
	No (-1), \newline
    Yes (+1) \\ \addlinespace 
Crimes & Did one of the two sides attack civilians, execute prisoners, or perpetrate other unlawful acts? & %
	Only Ukraine (-1), \newline
     Only Russia (+1) \\ %
\bottomrule
\end{tabular}
\label{tab:issues}
\end{center}
\end{table}

\section{Results}
\label{sec:results}

\subsection{Structurally polarized users fill the debate space (RQ1)}
\label{sec:polarization_over_time}

\begin{figure}[t]
    \centering
    \includegraphics[width=0.8\linewidth]{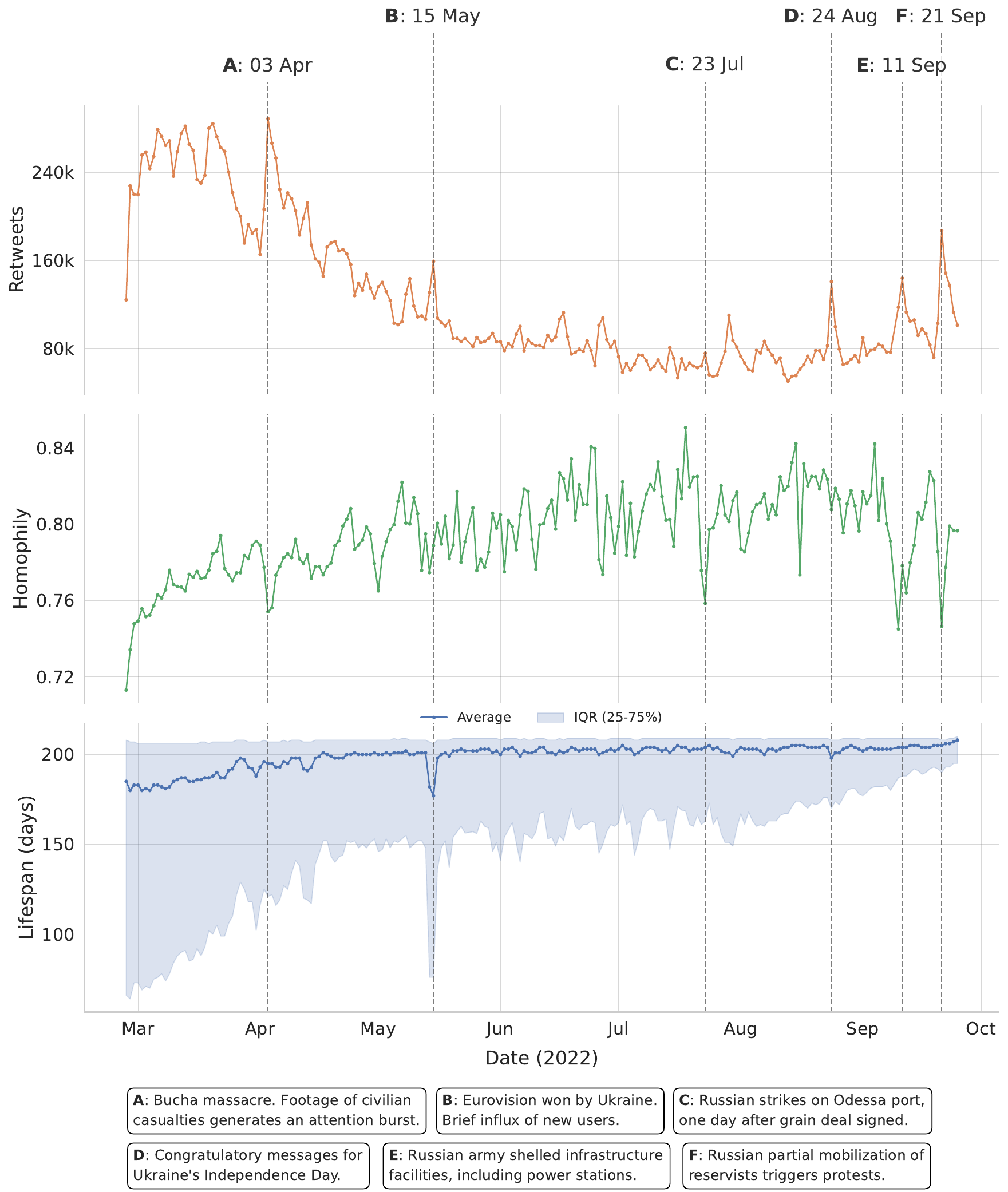}
    \caption{Dynamics of the Twitter debate on the Ukraine war over time. 
    (Top) The total daily number of retweets (red), a proxy for discussion volume, peaks at the invasion's start and subsequently declines. 
    (Middle) The mean homophily (green) of retweets steadily increases, indicating that the conversation network becomes more structurally polarized. 
    (Bottom) The median active user lifespan (dark blue line) and its interquartile range (shaded area) also rise, showing that the debate is increasingly sustained by a persistent ``hard core'' of long-term users.
    The vertical dashed lines labeled with letters [A--F] and specific dates represent notable events.}
    \label{fig:debate_evolution} %
\end{figure}

We first analyze the longitudinal dynamics of the debate by examining three key metrics over time, as shown in \Cref{fig:debate_evolution}.
The top panel of \Cref{fig:debate_evolution} shows the \emph{total number of retweets} in our dataset, a proxy for discussion volume.
This metric peaks during the last week of February 2022, immediately following the invasion, but then enters a steady decline, indicating waning public engagement on Twitter.
The middle panel plots the average daily \emph{homophily}---the fraction of retweets where users repost content from their own community (see
\Cref{sec:polarizingtweets}).
Homophily shows a clear upward trend,
rising from approximately \num{71}\% to over \num{82}\%.
This trend demonstrates that the debate space becomes progressively more fractured: information flows concentrate within communities rather than across them.
This pattern is typical of \emph{structural polarization}~\citep{interian2023network,salloum2022separating,preti2026dsp}, %
which often appears in online discussions on controversial topics~\citep{garimella2017effect,garimella2018quantifying} %
and is a prerequisite for the formation of echo chambers~\citep{garimella2018political,cinelli2021echo}.

The reason for this increasing structural polarization, however, is not necessarily a change in the retweeting behavior of individual users over time.
In other words, users do not necessarily become more homophilic or polarized as the war unfolds.
Instead, the data points toward a selection effect driven by a shift in the active user population.
The bottom panel of \Cref{fig:debate_evolution} shows the median \emph{lifespan} of users active on any given day, measured as the number of days since their first post in the dataset.
This median lifespan steadily increases as the discussion progresses: less engaged users drop out, and the users who remain active for long are, on average, more homophilic, more engaged with the discussion, and thus arguably those with the most strongly held convictions.
This mechanism runs contrary to the common narrative of social media platforms actively radicalizing their users over time.
Rather, the platform's environment appears to select for users whose interaction patterns are already more insular.
A few events, such as the release of video footage of the Bucha massacre on 3 April, drive bursts of attention and reduced polarization, with casual users re-joining the debate.

\subsection{Temporal activity coincides with community issue stances (RQ2)} %
\label{sec:opinion_polarization}

\begin{figure}[tbp]
    \centering
    \includegraphics[width=\textwidth]{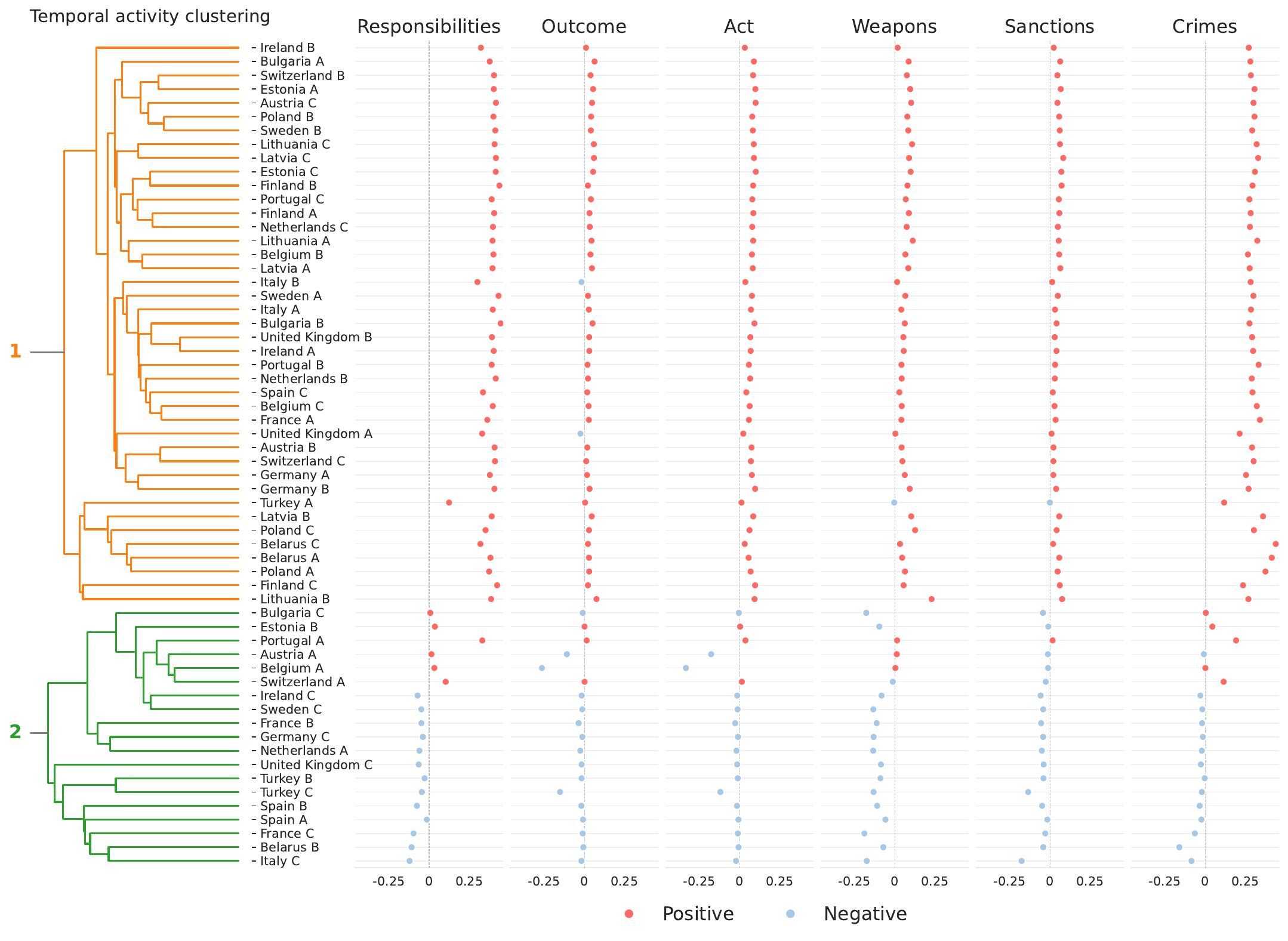}
    \caption{Community-level stances and temporal clustering. 
    On the left, the dendrogram displays the hierarchical clustering of communities, based on a similarity measure derived from the peaks in the time series of tweet volume for each community. 
    On the right, the plot represents the average stance score of retweeted labeled tweets for the top three communities in each country. Each point corresponds to a community, with the tweet's score computed as the average score assigned by four annotators, and its contribution is weighted by the total volume of its retweets.
    Positive scores are marked with a red dot, negative scores with a blue one.
    The column axes are the ones described in \Cref{tab:issues}.
    }
    \label{fig:communities_issues}
\end{figure}

To check whether such fragmentation is a sign of opinion polarization, we first consider the temporal attention dynamics of each community.
We thus measure which communities synchronize their attention, within or across countries.
The dendrogram in \Cref{fig:communities_issues} shows the hierarchical clustering of the top three communities by number of users in each country by comparing their daily posting dynamics.
We find that the communities can be grouped into two distinct clusters, meaning that one cluster's members post at different times than those in the other cluster. 
All countries, with the notable exception of Lithuania, Latvia, Finland, and Poland, have one of their top communities in one cluster and another in the other cluster, showing that significant debate on these issues is present within most European countries.
Each of these two clusters also exhibits notable differences among the communities inside it, corresponding to deeper levels of the dendrogram.

\begin{table}[!h] %
\caption{Issues, number of retweets labeled about the issue, percentage of all labeled retweets having that label, and two example posts (translated to English, where necessary, and all rephrased to preserve anonymity). Note that examples may apply to several issues.
}
\begin{center}
\begin{tabular}{l r r l}
\toprule
Title & Retweets & \% Labeled & \\
\midrule

\textbf{Responsibilities} & \num{7776450} & 61\% & (95\% Russia / 4\% NATO) \\
\multicolumn{4}{p{0.95\linewidth}}{  \hspace{0.4cm} 
This is not simply a conflict between Russia and Ukraine; it's a confrontation between NATO and Russia, where Ukraine serves as a proxy and its people bear the consequences of NATO’s expansion and hostility. } \\
\multicolumn{4}{p{0.95\linewidth}}{ \hspace{0.4cm} To those who ask, “Maybe Crimea and Donbas should be surrendered in return for peace?” — before posing that question, tell me which exact region of your own country you would be ready to hand over if Putin began bombing your cities. } \\ \addlinespace

\textbf{Outcome} & \num{1677982} & 13\% & (54\% Justice / 46\% Peace) \Tf\\
\multicolumn{4}{p{0.95\linewidth}}{ \hspace{0.4cm}
The brutal assault on \#Ukraine shows no sign of stopping—a tragic and irrational bloodshed in which destruction and crimes occur day after day. Nothing can excuse this. I urge the global community to commit to bring this abhorrent conflict to an end. \#Peace} \\
\multicolumn{4}{p{0.95\linewidth}}{ \hspace{0.4cm} 
Those responsible for crimes must be identified as criminals, prosecuted, and punished. The images from \#Bucha shatter the notion that a compromise should be pursued at any cost. In reality, Ukraine's defenders urgently require three things above all else: weapons, weapons, and still more weapons. \#StandWithUkraine } \\ \addlinespace

\textbf{Act} & \num{1418093} & 11\% & (88\% Negotiate / 12\% Fight) \Tf\\
\multicolumn{4}{p{0.95\linewidth}}{ \hspace{0.4cm} 
I urge international organizations and every nation to genuinely back Turkey’s efforts to achieve a lasting peace in Ukraine. \#UNGA} \\
\multicolumn{4}{p{0.95\linewidth}}{ \hspace{0.4cm} My Finnish volunteer comrades fighting alongside Ukraine. What do we truly live and die for? Is it to gather wealth and build comfortable careers, or to stand for others and for the principles we firmly believe in? \#Ukraine \#UkraineRussiaWar \#StandUpForUkraine [URL] } \\ \addlinespace

\textbf{Weapons} & \num{1369012} & 11\% & (48\% Yes / 52\% No) \Tf\\
\multicolumn{4}{p{0.95\linewidth}}{ \hspace{0.4cm} 
So far, Estonia’s humanitarian and military assistance to Ukraine totals 0.8\% of our GDP. But everyone needs to do more. If Ukraine hasn’t won, then we haven’t done enough. Deeds matter more than words. [URL]} \\
\multicolumn{4}{p{0.95\linewidth}}{ \hspace{0.4cm} 
Alert: Pascal Boniface, head of IRIS, claims that the massive financial and military assistance sent to Ukraine is “already being siphoned off by mafia networks” in a country he describes as “more corrupt than Yugoslavia,” adding that some of the supplied weapons are reportedly ending up in other countries. [URL]} \\ \addlinespace

\textbf{Sanctions} & \num{988155} & 8\% & (51\% Yes / 49\% No) \Tf\\
\multicolumn{4}{p{0.95\linewidth}}{ \hspace{0.4cm} 
The EU’s economic campaign against Russia does little to assist innocent Ukrainians suffering from this unlawful war. Moscow remains unaffected, while people across the EU endure soaring inflation, surging energy prices, and an unprecedented drop in living standards. This is irrational. [URL]} \\
\multicolumn{4}{p{0.95\linewidth}}{ \hspace{0.4cm} 
NATO Secretary General @jensstoltenberg told Members of the European Parliament: ``The cost we bear in the EU and in \#NATO can be calculated in financial terms. The cost Ukrainians face is counted in lives lost each day. So we must stop complaining, step up our efforts, and deliver the support needed. That’s it.'' \#StandWithUkraine }\\ \addlinespace

\textbf{Crimes} & \num{3938184} & 31\% & (96\% Russia / 4\% Ukraine) \Tf\\
\multicolumn{4}{p{0.95\linewidth}}{ \hspace{0.4cm}
Thread: In Ukraine, hundreds of people have reportedly been persecuted for various reasons by paramilitary units and the National Guard. Graphic footage shows torture, abuse, and humiliation, including of women and children. [URL]}\\
\multicolumn{4}{p{0.95\linewidth}}{ \hspace{0.4cm} 
I’m only beginning to grasp the magnitude and scope of the atrocities committed by Russian forces in the Kyiv region over the past two weeks. And it’s not limited to Irpin, Bucha, Hostomel, Borodyanka, and the like. Scores of smaller villages were entirely terrorized, isolated, and civilians were executed.} \\

\bottomrule
\end{tabular}
\label{tab:issue_stats_examples}
\end{center}
\end{table}

We then investigate whether this divergence in attention is accompanied by a divergence in their stances related to the conflict (RQ2).
Using the labels introduced in \Cref{sec:axis-labeling},
\Cref{tab:issue_stats_examples} shows statistics of the labeled dataset, including the numbers and percentages of retweets labeled as relevant to each issue. 
We also show two example tweets annotated as relevant to the issue (rephrased, to limit the exposure of the original poster).
The Responsibilities and Crimes axes are most present in our data, whereas those concerning the Actions, Weapons, and Sanctions, least; though even the least prominent issue (Sanctions) gathers nearly a million retweets.
On the issues, different stances prevail: an overwhelming majority emphasizes Russian responsibilities in the war and in committing war crimes, but at the same time, a majority prefers negotiating over supporting the fight.
The tweets are split in half on weapons and sanctions.

\Cref{fig:communities_issues} shows the average score (as defined in \Cref{sec:axis-labeling}) by community, for each of the six issues, ordered by the temporal clustering in the dendrogram. 
We find that the two clusters of communities have stances of opposing signs on most issues. 
In fact, the distribution of scores is different between the two clusters for all issues at $p<10^{-4}$ using the Mann-Whitney $U$ test. %
Thus, we find that these two clusters are distinct both in terms of their temporal attention and in their expressions on key issues, showing evidence of polarization.

The most polarizing issues are about which side is held responsible for starting the conflict, which has committed war crimes, and whether weapons should be sent to Ukraine.
The first cluster views Russia as the aggressor and having engaged in war crimes, and is more positive on sending weapons to Ukraine and imposing sanctions on Russia.
To collect this set of related stances, we use the term ``hawkish''~\cite{krastev_leonard_2022_peace}, as it seeks to punish Russia, support Ukraine, is more favorable to sending weapons, and indicates justice as the main goal rather than peace.
The second cluster, on the other hand, emphasizes the responsibility of NATO and the EU for the conflict, is less likely to accuse Russia of war crimes, and is more opposed to the supply of weapons to Ukraine and the imposition of sanctions on Russia. 
We call these stances ``doveish'' \cite{krastev_leonard_2022_peace}, as they seek to avoid escalation and indicate peace as the more important goal rather than justice.
We stress that these two clusters contain multiple communities, each with complex stances on these issues.
However, we find that there is a synchronization between communities in each cluster, both in terms of temporal activity and stance. 
As we will see next, this broad distinction will prove insightful for our subsequent analyses.

\subsection{Synchronization across Europe differs by stance (RQ3)}
\label{sec:Opinion_expression_synchronizes_across_Europe}

\begin{figure}[p]
    \centering
    \includegraphics[height=0.91\textheight, width=\linewidth, keepaspectratio]{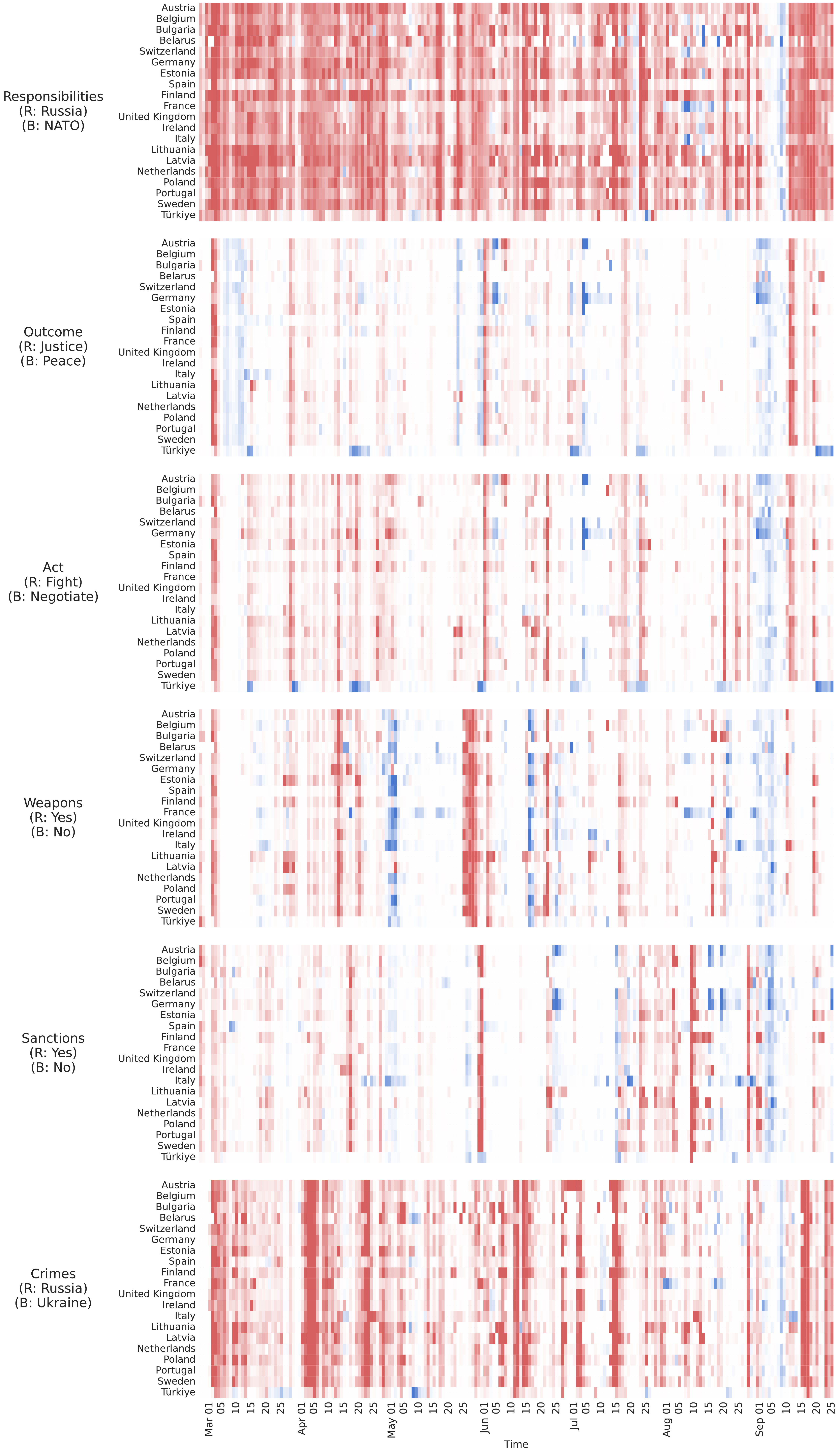}
    \caption{Temporal dynamics of the political positions expressed by the polarizing retweet activity in each country.
    The color represents the average score among the labeled tweets that were retweeted on a given day by a user geolocalized in a given country.
    In the title of each panel, we report first the stance for blue, and then the one for red.
    }
    \label{fig:rwb}
\end{figure}

To understand whether there is side-specific synchronization across countries on each of the issues, indicative of a pan-European attention dynamic, we combine the issue stance and temporal activity.
\Cref{fig:rwb} shows the country-wide temporal dynamics of the aggregated stance scores for each of the six issues. 
To compute the country-wise stance score for a day, we average the scores of the tweets posted by all communities of that country within the day (including the retweets). 
As the exact scores are not informative, we show the strength of the expressed stance using color, with white signifying 0 or neutral. 

First, different issues present different majoritarian stances, as indicated by \Cref{tab:issue_stats_examples}, with the attribution of Russian responsibility for the war being the most clear majority.
Then, the figure shows that, on most of the issues, the spikes in activity often happen at the same time (vertical lines). Visually, the hawkish stances (in red) is more synchronized than the doveish (in blue).
A remarkable exception is Turkey, which often has a distinct opinion dynamic from other countries in the dataset. 
There are also country-specific examples of countries displaying stances diverging from those of the other countries.
For instance, Italy shows more opposition to the sanctions throughout the timeline, as well as Austria, Switzerland, and Germany, whose activity shows anti-sanctions spikes in late June and late August. 
Germany, for instance, had a lively discussion around the possible re-opening of the NordStream2 gas pipeline, which would lift the gas embargo on Russia and supposedly lower the gas prices in Europe.\footnote{\url{https://www.welt.de/politik/deutschland/article240560467/Wolfgang-Kubicki-fuer-Inbetriebnahme-von-Nord-Stream-2-FDP-empoert.html}} 
Similarly, France stands out with doveish activity on the weapons issue.

\begin{figure}[t]
    \centering
    \includegraphics[width=\linewidth]{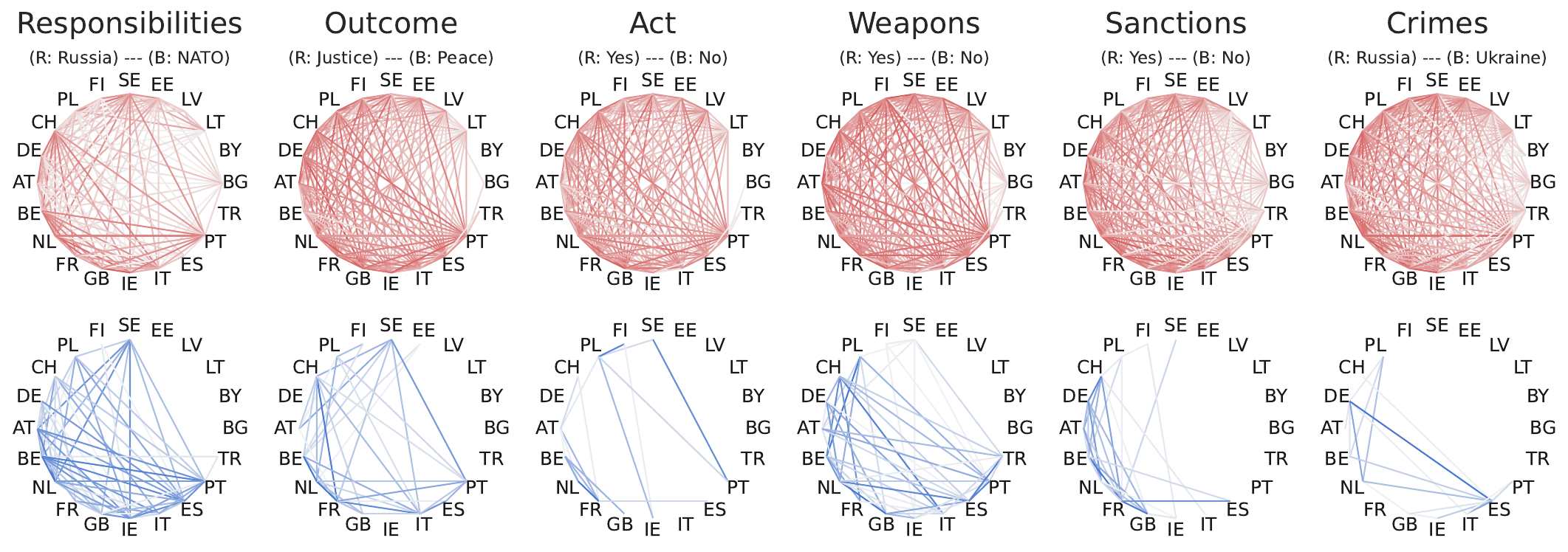 }
    \caption{Pairwise Spearman correlation networks of Hawkish and Doveish sides across countries and axes (meaning of each axis is described in \Cref{tab:issues}). Edges represent the Spearman rank correlation between country time series.  Only statistically significant connections (p < 0.05) with a correlation coefficient $\rho \ge 0.65$ are shown.}
    \label{fig:correlation-networks}
\end{figure}

To check more rigorously whether there are countries that synchronize on different issues significantly, \Cref{fig:correlation-networks} shows the Spearman correlation of the retweet volume of each stance.
The two clusters also differ in how synchronized they are, with the hawkish side more in synch across Europe, while the doveish side is more idiosyncratic to individual countries.
The hawkish (red) side of the issue is very internationally synchronized across all issues, and a bit less so around the responsibility one.
Conversely, the doveish (blue) side is much less synchronized, with the responsibility issue being most synchronous.
One possible explanation for this difference is that the hawkish side reflects a more pan-European opinion movement that supports Ukraine and defines itself in antagonism with Russia~\cite{makarychev2014russiaAndEU}, while the doveish side is more tied to internal, national debates.

\begin{figure}[htp]
    \centering
    \includegraphics[width=0.6\textwidth]{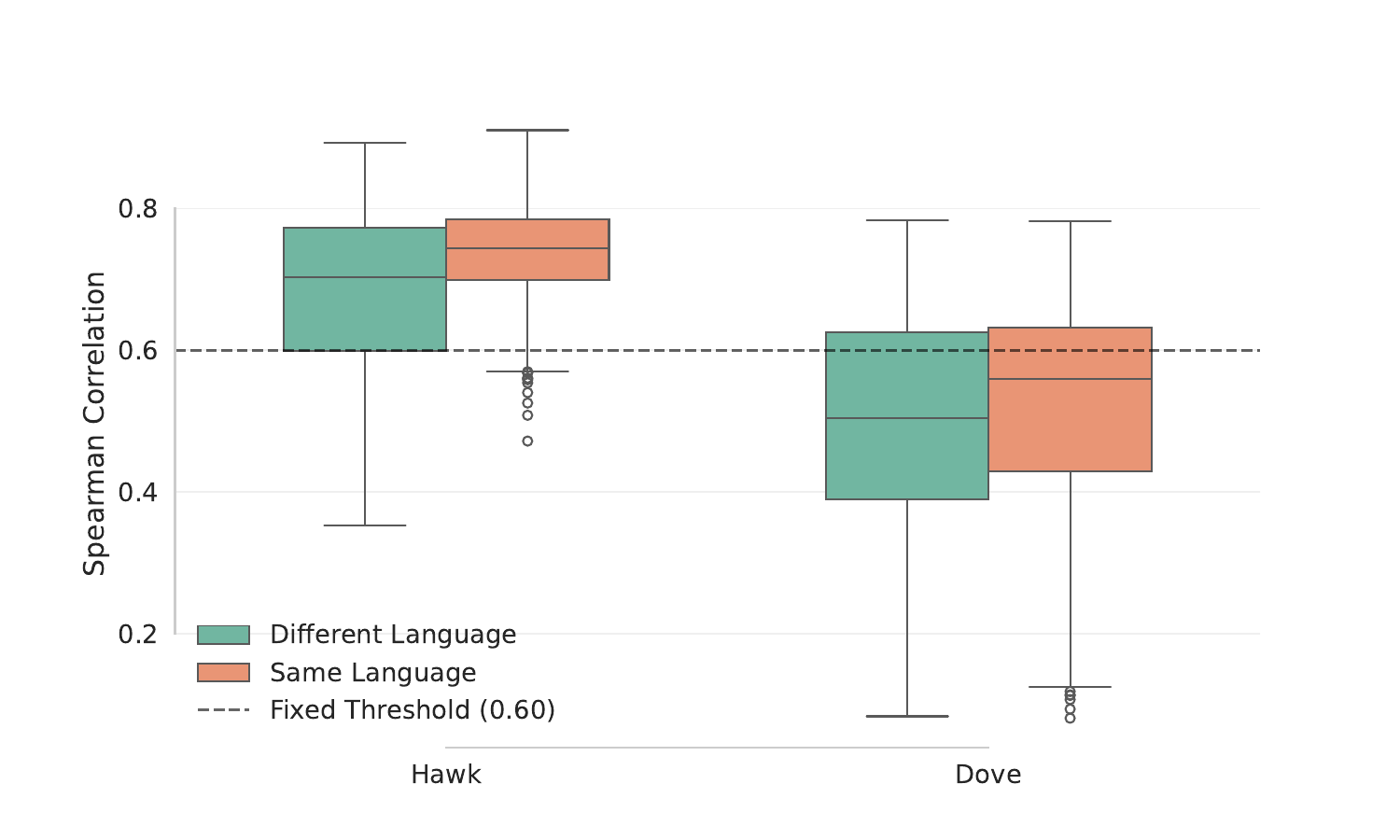}
    \caption{Distribution of pairwise Spearman correlations for country pairs sharing the same data language (green) versus different languages (orange). ``Same Language'' pairs show statistically higher correlations for both Hawkish and Doveish (Mann-Whitney $U$, $p < 0.05$). This result suggests that a shared language is a statistically significant, yet modest, driver of synchronization.}
    \label{fig:correlation_language_SPEARMAN}
\end{figure}

In the figure, we find the echoes of the above observations, such as the synchrony between Austria, Switzerland, and Germany about the sanctions. 
We also check whether the countries with the same selected language are more likely to correlate than those with a different ones: \Cref{fig:correlation_language_SPEARMAN} shows that, indeed, there is some linguistic affinity (countries with the same selected language have higher correlations, significant at $p<0.05$ using Mann-Whitney $U$ Test), but the difference between the issue sides (hawkish vs.~doveish) is more prominent than the one explained by language alone. 
It is crucial to note that such high degrees of pan-European temporal alignment may reflect not only organic shared attention but also the presence of highly mobilized activist networks or coordinated information operations, a known dynamic in the digital ecosystem of this conflict~\citep{cinelli2022coordinated}.

\subsection{Opposing sides focus on the same events only when debating specific actions (RQ4)} %
\label{sec:intrinsic_vs_extrinsic_focus}

To answer our final research question, we seek to understand whether the attention peaks we observe for each stance (the vertical lines of \Cref{fig:rwb}) can be attributed to external events, or follow an internal dynamic purely driven by the virality of individual tweets.
To do so, we consider each stance in each country, and compute daily totals for the scores of tweets supporting that stance. %
Then, we consider the top 3 peaks of activity for each stance and country.
To evaluate whether these peaks are endogenous or exogenous, we consider among all the possible pairs of peaks across opposing stances those separated by at most 7 days.
We then pick for each of those pairs of close-by peaks, the most important tweet for each side, computed by multiplying the number of retweets a post has received on that day by its opinion score, and selecting the tweet with the highest such score on that day for each side.
This process results in 121 pairs of tweets that were posted in a given country around the same time, but which have opposite stances on a specific issue.

All of the authors of this study then annotate all of the pairs of tweets for (i) whether each tweet concerns an event, and (ii) whether the two opposing tweets mention the same event.
Following \citet{mourelatos1978events}, we consider an event as a distinct occurrence that happened within a small period of time, and is associated with a specific place or persons  
(for instance, a specific military action, a speech by a prominent politician, or a release of official statements).

\begin{figure}[t]
    \centering
    \includegraphics[width=0.5\linewidth]{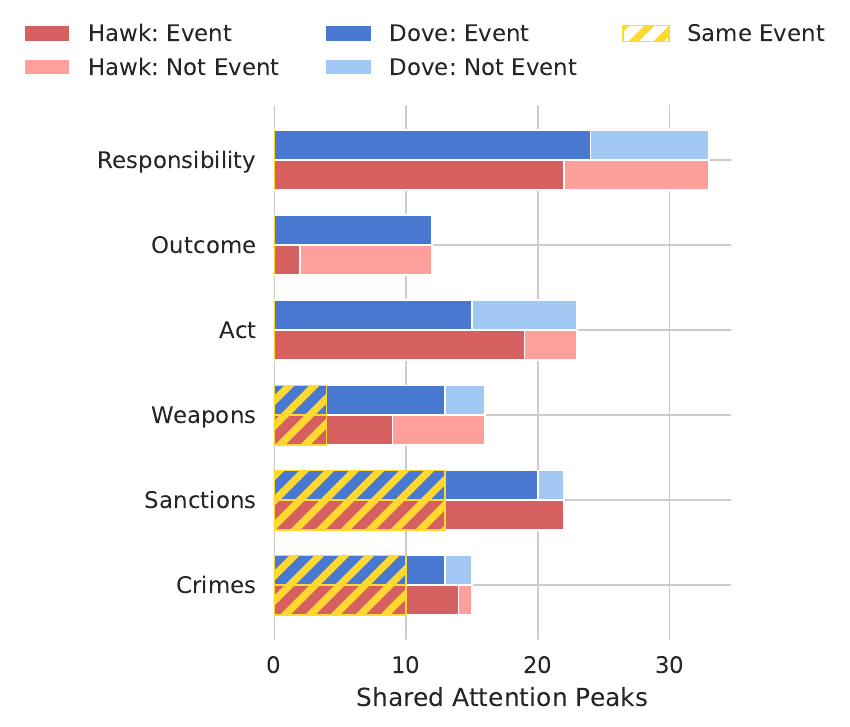}
    \caption{Number of tweets corresponding to the shared attention peaks by each side (red/blue), whether they are related to an event (dark/light), and whether they are related to the same shared event (yellow pattern). The meaning of each axis is described in \Cref{tab:issues}.}
    \label{fig:events_bar_plot}
\end{figure}

\Cref{fig:events_bar_plot} shows the number of tweets by the dovish (blue) and hawkish (red) sides of each issue axis (as defined in \Cref{tab:issues}) that are related to an event, as well as those instances where the two sides speak about the same event (in yellow).
On average, the hawkish side talks about an event 73\% of the time, and the doveish side 80\%. 
In general, both sides mention ongoing events, except the pro-justice, hawkish side of the outcome axis, and both sides of the responsibility axis.
In those exceptions, attention is often driven endogenously by viral tweets.
On all other axes, the top peaks are related to exogenous events from the external world.

However, whether their discussion converges on the same event strongly depends on the issue at hand.
This finding is surprising since, by design, the peaks in volume happen around the same time.
Moreover, we observe a sharp division between the two types of issues defined in \Cref{sec:interpretive-pragmatist}.
For the three interpretive issues, the conversations on the two opposing sides focus on divergent events and meanings.
Instead, attention peaks from both sides of pragmatist issues are catalyzed by high-profile events.
In other words, on issues involving the responsibility for the war, its outcome, and how Europe should behave, the two sides foreground different parts of the world.
Instead, issues concerning specific actions related to the supply of weapons to Ukraine and the sanctions policies, or specific claims of war crimes, create a focused, antagonistic debate around these concrete events.
Inspecting these tweets, it emerges that both evidence of war crimes (notably the Bucha massacre) and public discussion in institutional bodies (such as the European Parliament) drive attention from both sides of those issues.

\section{Limitations}
\label{sec:limitations}

Several limitations should be considered when interpreting the findings of this study.
It relies on retweets as a primary signal of attention and community alignment, which is a heterogeneous communicative practice:
it can reflect endorsement, but also amplification, ironic critique, or strategic visibility~\citep{boyd2010tweet}.
The networks reconstructed here, therefore, capture patterns of amplification and coordinated visibility rather than stable opinions or beliefs.
Furthermore, the network analysis excludes original tweets that were not retweeted: our analysis focuses on the most visible messages at the expense of lower-visibility conversational exchanges. %

The labeled subset used for stance annotation and event analysis introduces a related limitation.
The annotators are also European-based researchers and their positionality may have introduced subtle interpretive tendencies.
The observed patterns of synchronized attention may also reflect coordinated inauthentic behavior rather than organic mobilization alone~\citep{geissler2023russian}.
Tightly aligned attention spikes are among the known signatures of such operations~\citep{cinelli2022coordinated}.
Because the present study does not attempt to detect bots or analyze coordination, the findings should be interpreted with this caveat in mind.
Nevertheless, even coordinated media activity are themselves analytically meaningful, as they form part of the observable information environment and contributes to the dynamics of attention, visibility, and perceived consensus.

Our results also depend on specific platform affordances.
Twitter's sociotechnical architecture is optimized for real-time, episodic engagement, driven by trending topics and breaking news~\citep{hermida2010twittering,bruns2015twitter}.
When debate concerns concrete actions or atrocities, the platform's affordances facilitate rapid circulation of high-profile events across networks. %
However, interpretive debates regarding historical responsibility or abstract outcomes lack these viral, episodic anchors.
While engagement optimization is a general feature of social media platforms, Twitter's specific event-binding mechanism relies on its architecture as a networked public,
where content travels through explicit social ties rather than algorithmic interest graphs~\citep{hermida2010twittering}.
On platforms organized around algorithmic interest curation, such as TikTok's, eventification may be structurally weaker~\citep{defranciscimorales2021no,gerbaudo2026tiktok}.
Whether our results apply to other platforms requires further investigation.

Finally, the analysis focuses on an exceptional geopolitical crisis.
Wartime conditions compress the range of legitimate public debate and produce elite-driven framing dynamics absent from routine political contexts~\citep{entman2014media}.

\section{Discussion}
\label{sec:discussion}

Our findings suggest that polarization in digital public spheres is multi-layered.
It is structural, in that network homophily increases as participation narrows.
It is substantive, in that communities express coherent and opposing stances across multiple axes.
It is spatial, in that most countries host both sides of the divide.
And it is conditional, in that the degree of shared referential frame depends strongly on the type of issue under discussion.

On the pragmatist issues, both clusters often focus on the same high-profile events.
They draw radically different normative conclusions, but they react to the same concrete occurrences---new evidence, documents, or institutional decisions.
When debate concerns tangible policy tools or legally defined events, external constraints impose a minimal shared referential anchor,
where disagreement concerns how to respond to a commonly recognized situation.
As in the Habermasian public sphere, even intense debate unfolds within a common horizon of reference~\citep{habermas1981reason,habermas1981lifeworld,habermas1992further}.
Yet,
opponents engage the same events as \emph{adversaries} contesting meaning, rather than as deliberators oriented toward consensus,
as in Mouffe's \emph{agonistic} public sphere~\citep{mouffe2000democratic}.

By contrast, on interpretive issues, the two sides rarely converge on the same events, but they refer to distinct episodes and causal chains.
In these instances, %
we observe \emph{affective publics}~\citep{papacharissi2014affective}, where communities synchronize around emotionally resonant narratives and not evidentiary evaluation, and in some cases \emph{counterpublics}~\citep{warner2002publics},
in the form of nationally-addressed counter-discourse that circulates in tension with pan-European, hawkish consensus~\cite{eu2022publicopinion}.
Data reveal an asymmetrical synchronization consistent with horizontal Europeanization of the debate: while scholars have long debated the existence of a transnational ``European public sphere''~\citep{risse2010community,koopmans2010europeanization}, our results suggest that transnationalization on social media is heavily dependent on political alignment.

Conditional publics are not a distinct type of public, but a relational property of public formation, whereby the degree of shared reference across opposing groups varies systematically with issue type.
They are characterized by a dual mode of existence:
They coalesce as affective publics or counterpublics on interpretive issues, each constructing its own universe of salient events and meanings, yet reconstitute themselves as an agonistic public sphere on pragmatist issues, where a shared referential anchor forces engagement with the same concrete occurrences.
This concept extends existing theories, which tend to treat public formations as stable, by foregrounding issue type as a structuring variable that determines whether opposing communities inhabit a shared or fractured reality.
Its necessity stems from real-world political debates that span both issue types simultaneously: a framework that treats publics as uniformly deliberative, affective, or oppositional cannot account for the within-debate variation we observe.
In this sense, the European debate on the war in Ukraine allows us to clearly see these conditional publics in action, contributing to public opinion formation amidst a major geopolitical crisis.

\bibliographystyle{agsm}
\bibliography{references}

\end{document}